\documentclass{aa} 
\usepackage{graphicx,natbib,url}
\bibpunct{(}{)}{;}{a}{}{,} 
\usepackage{txfonts}
\newcommand{\gb}{\ensuremath{\gamma_{\rm j}\beta_{\rm j}}}

\newcommand{\urlfn}[1]{\footnote{\url{#1}}}
\newcommand{\mybf}{\bf}
\renewcommand{\mybf}{}
\begin{document}
\title{Jet-lag in Sgr A*: What size and timing measurements tell us about the central black hole in the Milky Way}
\author{Heino Falcke\inst{1,2} \and Sera Markoff\inst{3} \and Geoffrey C. Bower\inst{4}}
\offprints{H.\ Falcke, \email{H.Falcke@astro.ru.nl}} \institute{Department of
  Astrophysics, Institute for Mathematics, Astrophysics and Particle Physics, Radboud University,
  P.O. Box 9010, 6500 GL Nijmegen, The Netherlands \and 
ASTRON, Oude
  Hoogeveensedijk 4, 7991 PD Dwingeloo, The Netherlands \and 
Astronomical Institute ``Anton Pannekoek'', University of
  Amsterdam, Kruislaan 403, 1098SJ Amsterdam, The Netherlands \and 
UC Berkeley, 601 Campbell Hall, Astronomy Department \& Radio Astronomy Lab,
  Berkeley, CA 94720, USA
}

\date{December 25, 2008} 

\abstract
{The black hole at the Galactic Center, Sgr A*, is the prototype of a 
  galactic nucleus at a very low level of activity. Its radio through
  submm-wave emission is known to come from a region close to the
  event horizon, however, the source of the emission is still under
  debate. A successful theory explaining the emission is 
  based on a relativistic jet model scaled down from powerful
  quasars.}
{ We want to test the predictive power of this established jet model
  against newly available measurements of wavelength-dependent time
  lags and the size-wavelength structure in Sgr A*.  }
{Using all available closure amplitude VLBI data from different
  groups, we again derived the intrinsic wavelength-dependent size of Sgr
  A*. This allowed us to calculate the expected frequency-dependent
  time lags of radio flares, assuming a range of in- and outflow
  velocities. Moreover, we calculated the time lags expected in
  the previously published pressure-driven jet model. The predicted
  lags are then compared to radio monitoring observations at 22, 43,
  and 350 GHz.}
{\mybf The combination of time lags and size measurements imply a
  mildly relativistic outflow with bulk outflow speeds of
  $\gamma\beta\simeq0.5-2$. The newly measured time lags are
  reproduced  well by the jet model without any major fine tuning. 
}
{\mybf The results further strengthen the case for the cm-to-mm wave
  radio emission in Sgr A* as coming from a mildly relativistic jet-like
  outflow. The combination of radio time lag and VLBI closure
  amplitude measurements is a powerful new tool for assessing the flow
  speed and direction in Sgr A*.  Future VLBI and time lag
  measurements over a range of wavelengths will
  reveal more information about Sgr A*, such as the existence of a jet
  nozzle, and measure the detailed velocity structure of a
  relativistic jet near its launching point for the first time.}

\authorrunning{Falcke, Markoff, Bower}
\titlerunning{Jet-lag in Sgr A*}

\maketitle

\section{Introduction}
\label{s:intro}
The Galactic center hosts by far the best constrained super-massive
black hole candidate: the compact radio source \object{Sgr A*}
\citep[see][for a review]{MeliaFalcke2001}. 
Its mass is believed to be around
$4\times10^6M_\odot$ based on stellar proper motion measurements
\citep{SchoedelOttGenzel2002,GhezSalimHornstein2005}. Linear polarization measurements indicate 
that it is extremely underfed, with an accretion rate of less than
$10^{-7}M_\odot/{\rm yr}$
\citep{Agol2000,BowerFalckeWright2005,MacquartBowerWright2006,MarroneMoranZhao2007}. The accretion rate and low radio flux put Sgr A* 
at the tail end of the local luminosity function
\citep{NagarFalckeWilson2005} of low-luminosity active galactic nuclei
(LLAGN). This makes Sgr A* an ideal laboratory to study supermassive
black hole physics in the quasi-quiescent state in which most galactic nuclei
exist today.

Sgr A* has been detected at radio \citep{BalickBrown1974} and now 
near-infrared \citep{GenzelOttEckart2003} and X-ray wavelengths
\citep{BaganoffBautzBrandt2001}.  The radio spectrum of the source is
variable, slightly inverted, and peaking in a submm-bump which
originates close to the event horizon
\citep{ZylkaMezgerLesch1992,FalckeGossMatsuo1998,FalckeMeliaAgol2000,MeliaFalcke2001,MiyazakiTsutsumiTsuboi2004,EckartBaganoffSchoedel2006}. 
The latter is of particular importance since it may eventually allow
imaging of the shadow cast by the event horizon
\citep{FalckeMeliaAgol2000,HuangCaiShen2007,BroderickLoeb2006}. However, until recently
no structural information was available for Sgr A*. At wavelengths
shorter than that of the submm-bump, the resolution of current telescopes is
insufficient and at long wavelengths, where high-resolution very long
baseline interferometry (VLBI) techniques can be used, the source
structure is blurred by interstellar scattering.

This ambiguity has led to a longstanding debate about the actual nature of the Sgr
A* emission. One class of models suggests that the radio through X-ray emission
is caused by accreting hot plasma flowing into the black hole 
\citep{Melia1992a,NarayanMahadevanGrindlay1998,QuataertGruzinov2000a}. 
On the other hand, it has been suggested that Sgr A* resembles the
compact radio cores of active galactic nuclei (AGN)  and therefore most of the
emission is associated with a (mildly) relativistic outflow or jet
\citep{ReynoldsMcKee1980,FalckeMannheimBiermann1993,FalckeBiermann1999,FalckeMarkoff2000,YuanMarkoffFalcke2002}.

Only recently have measurements of the intrinsic size of Sgr A* become
available
\citep{BowerFalckeHerrnstein2004,ShenLoLiang2005,DoelemanWeintroubRogers2008},
providing crucial new input. The new intrinsic size measurements
agree well with the predictions of the traditional jet model
\citep{BowerFalckeHerrnstein2004,MarkoffBowerFalcke2007}, however, a
direct confirmation of an outflow is still lacking.

Clearly, additional information is required to determine the speed and direction of the flow responsible for the emission in Sgr A*. Such additional information has now become available with the first reliable time lag measurements of radio outbursts at different wavelengths \citep{Yusef-ZadehRobertsWardle2006,Yusef-ZadehWardleHeinke2008}.{\mybf These   observations show that high radio frequencies lead the lower radio frequencies by some   20 minutes around 43 GHz. Because the radio emission is considered to be optically   thick due to its flat-to-inverted spectrum, and the synchrotron loss timescale is much   longer, the radio flux variations are tracing actual adiabatic expansion or contraction   of the emitting plasma. This scenario is in marked contrast to observations in the   optically thin part of the spectrum at near-infrared (NIR) and X-ray bands, where the cooling time scales are faster than adiabatic.  At these higher frequencies,  observations  \citep{MarroneBaganoffMorris2008,Dodds-EdenBartkoEisenhauer2008} show a near   simultaneity between NIR and X-ray flares within minutes and a delay between X-ray/NIR   with respect to the radio emission on ther order of hours. The expectation therefore is that   radio timing observations trace bulk plasma properties, while X-ray/NIR variability is   dominated by heating and cooling of particle energy distributions in the plasma. Which   physical parameters determine a potential lag between X-rays/NIR and radio/submm   \citep{MarroneBaganoffMorris2008}, is not immediately obvious.}

In this paper we focus on the radio time lag data and size
measurements to obtain information on the plasma flow speed. To do
this we re-derive the intrinsic size of Sgr A* by combining all
existing VLBI data in Sec.~\ref{s:vlbisize}, thereby resolving some of
the apparent discrepancies between the results of different groups in
the literature. We then compute the predicted time lags for various
inflow/outflow speeds in Sec.~\ref{s:timedelay} and present time lag
predictions of the canonical jet model in Sec.~\ref{s:jetmodel}. Here
we also present the only analytical velocity profile of a
pressure-driven jet in a closed form. The predictions are then
compared with the data {\mybf under the assumption that the region
  causing the variability roughly follows a similar size-frequency
  relation as seen by VLBI, tracing the bulk of the plasma}. Our main
conclusions are then summarized and discussed in
Sec.~\ref{s:conclusions}.

\section{Size and time lag data in Sgr A*}
\subsection{VLBI size of Sgr A*}
\label{s:vlbisize}
The radio size of Sgr A* is extremely difficult to determine for several reasons. The radio source itself is very compact and hence VLBI techniques have to be used, where radio telescopes with separations of several thousand kilometers are combined to obtain interferometric information of the source structure. However, the major high-frequency VLBI telescopes are in the Northern hemisphere, making Sgr A* a low-elevation source which is difficult to calibrate. Moreover, the source is located in the Galactic center behind a large scattering screen that broadens the intrinsic source size significantly at long wavelengths. To escape scattering effects requires observing at shorter wavelengths, which are even more difficult to calibrate.  Hence, the breakthrough for the detection of the intrinsic size \citep{BowerFalckeHerrnstein2004} came via the introduction of closure amplitude analysis \citep{DoelemanShenRogers2001}, a method relatively insensitive to common station-based calibration errors.

Closure amplitudes provide good means to measure the source size with
very high accuracy, especially if the source structure is
simple. Since the broadening of the source structure by scattering
follows a $\lambda^2$ law \citep{DaviesWalshBooth1976,vanLangeveldeFrailCordes1992,LoShenZhao1998,BowerFalckeHerrnstein2004}
the actual source size $\phi_{\rm Sgr\,A*}$ is given by
\begin{equation}
\label{e:quadrature}
\phi_{\rm Sgr\,A*}=\sqrt{\phi_{\rm obs}^2-\phi_{\rm scatt}^2},
\end{equation}
where $\phi_{\rm obs}$ and $\phi_{\rm scatt}$ are the actually
observed and the expected scattering size respectively. $\phi_{\rm
  scatt}$ can be obtained by measuring the source size at long
wavelengths, where the intrinsic size is negligible, and extrapolating
with a $\lambda^2$-dependence to shorter wavelengths. The validity of
this extrapolation of the $\lambda^2$-law has been discussed by
\citet{BowerFalckeHerrnstein2004} and demonstrated using the measured
Gaussianity of the scattered image.

In the following we employ this procedure using all currently
available data, and discuss the origin of apparently conflicting results for
the wavelength-size structure of Sgr A*.

There are currently only four papers which contain reliable major {\it
  and} minor axes for Sgr A*: \citet{BowerFalckeHerrnstein2004} for
$\lambda$6 cm to $\lambda$7 mm data, \citet{BowerGossFalcke2006} for
$\lambda$24 cm to $\lambda$17 cm data, and \citet{ShenLoLiang2005} and
\citet{Shen2006} for $\lambda$3 mm and $\lambda$7 mm data. Sizes at
wavelengths longer than 20 cm are from VLBA closure-amplitude
measurements and at longer wavelengths high-quality VLA data is
available.  There is also a closure amplitude size at $\lambda$3 mm
from \citet{DoelemanShenRogers2001}, however, that was only reliably
obtained for a circular Gaussian source and has been superseded by
\citet{Shen2006}, who fit an elliptical Gaussian. In addition there
was one older measurement at $\lambda$1.3 mm, by
\citet{KrichbaumGrahamWitzel1998}, based on a single baseline
detection. {\mybf The latter has now been superceded by a more recent
  detection by \citet{DoelemanWeintroubRogers2008}, based on three
  baselines and higher signal-to-noise ratio. The $\lambda$1.3 mm
  observations yield the smallest sizes and the largest excursion from
  the scattering law. We therefore include these data points in our
  analysis for completeness, even though they do not represent a
  closure amplitude measurement and one cannot distinguish between
  major and minor axis. Their inclusion, however,
  does not change our results significantly.}

\begin{figure}
\resizebox{\hsize}{!}{\includegraphics{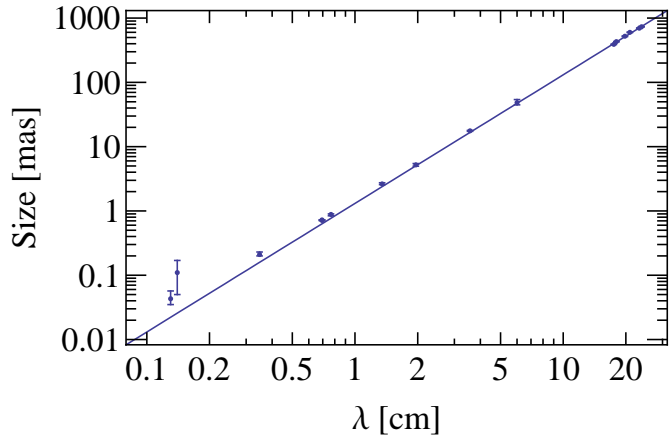}}
\caption{\label{f:sgr-size-vs-lambda}Measured radio source size (major axis) of Sgr A* as function of observing wavelength in centimeters.}
\end{figure}

Figure \ref{f:sgr-size-vs-lambda} shows the size data of Sgr\,A* as
function of wavelength together with the scattering law from
\citet{BowerGossFalcke2006}, $\phi_{\rm scatt}=(1.31\pm0.02)\,{\rm
mas}\;(\lambda/{\rm cm})^2$. One can clearly see how the overall size
of Sgr A* follows the $\lambda^2$-law closely at long wavelengths.

The next step is to subtract the scattering law in quadrature from the
observed size according to Equation~\ref{e:quadrature}. For this, the
exact normalization of the scattering law is vitally important. The
normalizations used by \citet{BowerGossFalcke2006} and
\citet{ShenLoLiang2005} differ only slightly. This has little impact on the
intrinsic source size at $\lambda$3 mm and $\lambda$7 mm, but markedly
affects the size at longer wavelengths. As
\citet{BowerGossFalcke2006} showed, this changes the
size-vs-wavelength relation (size
$\propto\lambda^{m}$). \citet{BowerGossFalcke2006} find power laws in
the range between $m=1.3$ and $m=1.7$.
\citep{ShenLoLiang2005}, who use only short-wavelength data, find $m=1.09$. 

The biggest problem, therefore, is the systematic uncertainty introduced
by the inclusion or non-inclusion of long-wavelength data sets. 
We investigate this uncertainty in the following discussion. Note however that the difference in
the scattering law primarily affects intrinsic sizes at long
wavelengths; short wavelength sizes are largely unaffected because the
contribution of the scattering angle to the observed size is much
less.

\subsection{Robustness of the Sgr A* size measurements}

\begin{figure}
\resizebox{\hsize}{!}{\includegraphics{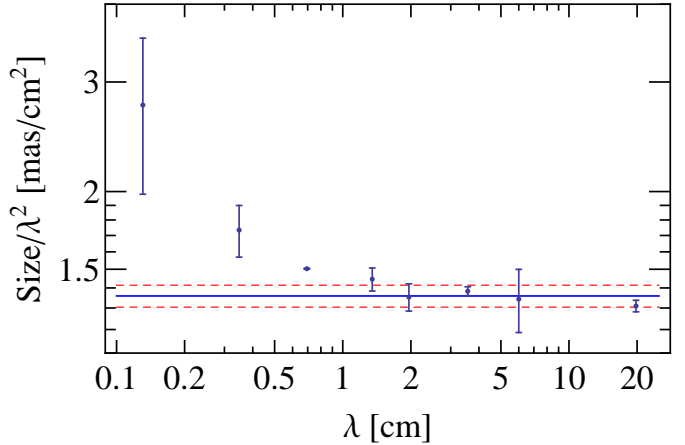}}
\caption{\label{f:sgr-normsize-vs-lambda}Measured radio source size (major axis) of Sgr A* divided by $\lambda^2$ as function of observing wavelength in cm. The solid line represents the average scattering law used here, where the dashed lines indicated the 3$\sigma$ limits found by randomly dropping one data point.}
\end{figure}

Figure~\ref{f:sgr-normsize-vs-lambda} shows the observed sizes divided
by $\lambda^2$. Here we have averaged the data for the various
observing bands, in order to avoid having the final fit be biased by the number of
observations in one band. For the averaging we divided the sizes by
$\lambda^2$ to take out the frequency dependence, and weighted them by
their error bars. This gives us one data point per band. In
particular, all 20 cm data from \citet{BowerGossFalcke2006} are
averaged into one point here. The error bars we show are the standard
deviations of the measurements in one band, where multiple
measurements were available. In principle this should be a more robust
measure of the error.

The non-homogeneous error distribution is problematic, but as it is 
a limit of the available observational data base, it cannot be overcome. The
three data points at $\lambda$7 mm, $\lambda$3.5 cm, and $\lambda$20
cm tend to dominate any fitting and a combined multi-parameter fit of
scattering-law and intrinsic size does not converge. Therefore it is
customary to only fit the scattering law to long-wavelength data. The
range of currently used scattering laws then depends exclusively on
which data to include. Any unknown systematic error at $\lambda$3.5 cm
or $\lambda$20 cm would drastically affect the result. To quantify the
robustness of the inferred sizes, we performed a series of
weighted fits to the data below $\lambda$1 cm, with one random data
point dropped. Doing this we find a range of possible scattering laws
(Fig.~\ref{f:sgr-normsize-vs-lambda}) given by
\begin{equation}
\phi_{\rm scatt}=(1.36\pm0.02)\,{\rm mas}\times(\lambda/{\rm cm})^2.
\end{equation}
This includes the best-fit scattering laws used by
\citet{BowerGossFalcke2006} and
\citet{ShenLoLiang2005} within 3 $\sigma$ limits, which have 
scaling factors of $1.31\pm0.02$ and $1.39\pm0.02$ respectively.

\begin{figure}
\resizebox{\hsize}{!}{\includegraphics{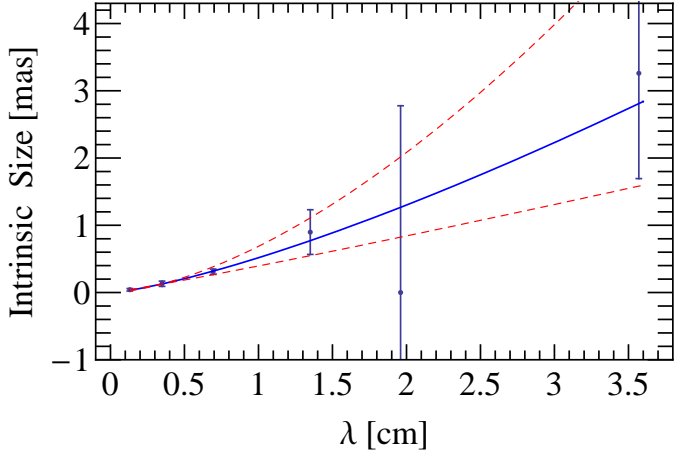}}
\caption{\label{f:sgr-intrinsicsize-vs-lambda}Intrinsic radio source size (major axis) of Sgr A* obtained by subtracting the scattering diameter in quadrature. The solid line represents the best fit powerlaw. The upper and lower dashed lines are the intrinsic sizes fitted by Bower et al. (2006) and Shen et al. (2005) respectively.}
\end{figure}

Subtraction of this scattering law in quadrature yields a slightly
revised intrinsic size as shown in
Figure~\ref{f:sgr-intrinsicsize-vs-lambda}.  The sizes at $\lambda$2
cm and $\lambda$3.5 cm are relatively sensitive to the scattering law
and therefore ``negative'' source sizes are possible within the
errors. Negative sizes are treated as lower limits around zero with
the respective error bars. We fit the error-weighted intrinsic source
size with a powerlaw function, yielding:

\begin{equation}\label{sgrsize}
\phi_{\rm Sgr\,A*}=(0.52\pm0.03)\,{\rm mas}\times\left({\lambda/{\rm cm}}\right)^{1.3\pm0.1}.
\end{equation}

Again, this is consistent with the previous results and will be used
in the following analysis.  We have further verified this result by
running a Monte Carlo simulation, excluding the 1.3 mm data, by
randomly varying the observed data and the scattering law within the
errors quoted here. To each of these trials we then fitted the
intrinsic size law and determined the slope parameter $m$. We find
that the distribution of $m$ is non-Gaussian, with a more extended
tail towards smaller values. The median, however, is again at
$m=1.44-0.19+0.16$, where the errors are the 25\% and 75\% quantiles,
respectively. These may be the more realistic error estimates than the
ones from the simple analytic fitting.

To improve on this result in the future, more and better
closure-amplitude size measurements need to be obtained at longer
wavelengths, especially at $\lambda$2 and $\lambda$6 cm.

In any case, the combined set of currently available data and the
error analysis confirm previous conclusions that there is a
wavelength-dependent photosphere in Sgr A* from a stratified
medium. As expected for optically thick synchrotron radiation, the
optical depth is indeed frequency dependent. This means that
observations of Sgr A* at two different radio wavelengths provide
information about two different spatial scales where the emission originates.

\subsection{Variability and time lags}
\label{s:timedelay}
In addition to the size measurement, we now have another new crucial parameter: the time lags between different wavelengths during small-scale variability outbursts. In the absence of direct imaging of source substructure, this provides the only means to determine flow or signal speeds in Sgr A*.

The overall variability of Sgr A* has been established for a long
time. The most comprehensive data sets stem from long-term monitoring
programs with the Green Bank Interferometer \citep{Falcke1999a} and
the VLA \citep{HerrnsteinZhaoBower2004} at cm wavelengths. The
reported rms variations of the radio spectrum are 2.5\%, 6\%, 16\%,
17\%, and 21\% at wavelengths of 13, 3.6, 2, 1.3, and 0.7 cm
respectively. \citet{MacquartBower2006} argue that most of the
variation at longer timescales (several days) and at long wavelengths
is due to interstellar scintillation. However, for time scales less
than four days the variations may be intrinsic with an rms of $\sim10\%$
for wavelengths 0.7-3 cm. Variability is also seen at mm and
sub-mm wavelengths \citep{ZylkaMezgerWard-Thompson1995,ZhaoYoungHerrnstein2003,MiyazakiTsutsumiTsuboi2004,MauerhanMorrisWalter2005,MarroneBaganoffMorris2008}
with yet larger rms variations and outbursts of a factor of several
over the quiescent level.


In most cases, where multiple wavelengths were observed, the time
coverage was not dense enough to find a reliable time lag between
two wavelengths, despite several attempts. Recently,
\citet{Yusef-ZadehRobertsWardle2006} and \citet{Yusef-ZadehWardleHeinke2008} published data obtained with the
VLA in fast switching mode allowing quasi-simultaneous high-time
resolution measurements of time variability in Sgr A* at two different
wavelengths. {\mybf They find a lag between $\lambda$1.3 and 0.7 cm on
the order of 20 minutes. Taking the weighted average of Table 1 in
\citet{Yusef-ZadehWardleHeinke2008} one finds a lag of $21\pm3$ minutes. 
Millimeter and submm-millimeter wave timing observations by the same
group are less significant, but seem to go in the same direction, with
a lag between 22 and 350 GHz of $65\pm+10-23$ minutes
\citep{Yusef-ZadehWardleHeinke2008}. }

The sign of the lag between 43 and 22 GHz ($\lambda\lambda$0.7 \& 1.3 cm) is such that the shorter wavelengths lead the longer ones. Keeping in mind that shorter wavelength emission originates at smaller size regions, this immediately implies that bursts propagate outwards from small to larger scales.  

Given that we know the projected size $s=\phi_{\rm Sgr\,A*}D_{\rm GC}$
of Sgr A* from observations -- $D_{\rm GC}$ is the Galactic center
distance $D=8$ kpc \citep{EisenhauerSchoedelGenzel2003} -- {\mybf the
time lag provides a straightforward estimate for the flow speed}. Using equation
Eq.~\ref{sgrsize} we find that the intrinsic size of Sgr~A* is
$\phi_1=0.73$ mas and $\phi_2=0.32$ mas or $s_1=8.8\times10^{13}$ cm
and $s_2=3.9\times10^{13}$ cm at $\lambda1.3$ cm and $\lambda7$ mm
respectively. Hence, $\Delta s$ is $\sim 27$ light minutes. Given a
time lag on the order of $\Delta \tau=20$ min the flow velocity is
$v=(s_1-s_2)/\Delta \tau=1.4 c$. Therefore, Sgr A* needs to a harbor
at least a mildly relativistic outflow, even if one allows for an
error of $\sim$50\%. Projection effects would tend to increase this
value even further.

\begin{figure}
\resizebox{\hsize}{!}{\includegraphics{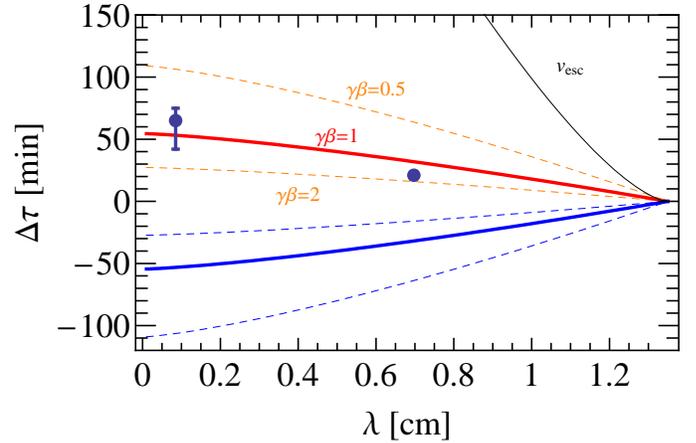}}
\caption{\label{f:sgr-time-delay-vs-lambda}Expected time lag of Sgr A* versus wavelength relative to $\lambda$1.3 cm (22 GHz) for the observed size-wavelength relation and a proper flow or signal speed of $\gamma\beta=1$ (red, solid line) or $\gamma\beta=0.5$ and 2 (orange, dashed). The data points are measured time lags from Yusef-Zadeh et al. (2008). The top black solid line shows the Newtonian time lag for an outflow just at the escape speed. Long lags above that line would correspond to gravitationally bound outflows.
}
\end{figure}

Alternatively, if one has a model for a flow speed $v(s)$ one can
predict the time lags with $\Delta\tau=(s_1/v(s_1) - s_2/v(s_2))$.
The easiest model is one with a constant flow speed $v(s)=$const. To
allow for relativistic speeds we write this as $v(s)=\gamma\beta c$,
where $\gamma=\sqrt{1-\beta^2}\,^{-1}$ and $\beta=v/c$. For this
subsection, we will ignore projection effects for the sake of
simplicity. The time lag then is $\Delta \tau=D_{\rm GC}\Delta
\phi_{\rm Sgr A*}/\gamma\beta c$. Figure
\ref{f:sgr-time-delay-vs-lambda} shows the time lags for proper speeds
$\gamma\beta$ in the range 0.5-2 c for the measured size-wavelength
relation.

We note that the source size in Sgr A* is close to linear with
wavelength, hence, for constant velocity one expects a linear increase
of the time lags with decreasing wavelength relative to a fixed
reference wavelength. For comparison, we also show the time lags if
the outflow would propagate always with the (Newtonian) escape speed
($\sqrt{2 GM/r}$) for a $3.6\times10^6M_\odot$ black hole. Here the
time lags would become longer and grow nonlinearly towards shorter
wavelengths, since the escape speed is significantly slower than the
speed of light. Figure \ref{f:sgr-time-delay-vs-lambda} shows that a
gravitationally bound flow would predict much longer time lags than
what is actually observed.

\subsection{Time lags in the jet model}
\label{s:jetmodel}
Given that the time lags suggest a mildly relativistic outflow, it
seems appropriate to investigate what an actual jet model
predicts. The basic jet model for Sgr A*
\citep{FalckeMannheimBiermann1993,FalckeMarkoff2000} naturally fits
the spectrum, properly predicted the low accretion rate of Sgr A* now
inferred from polarization, and also was able to explain the VLBI size
\citep{MarkoffBowerFalcke2007}. The only major property that could not
be tested so far is in fact the flow speed.

So far, the underlying assumption for the jet model has been that
Sgr A* is not a strongly relativistic outflow. Energetically
this is an optimal solution in terms of the ratio between total jet
power and emitted synchrotron radiation
\citep{FalckeMannheimBiermann1993}.  On the other hand, the sound
speed for a relativistic plasma as well as the escape speed from the
black hole are on the order of $\sim0.5c$, which sets a lower bound for a
supersonic jet in Sgr A*.

{\mybf In the standard \citet{BlandfordKonigl1979} model for the
flat-spectrum radio emission of compact jets, a constant velocity is
assumed and introduced as a free parameter. \citet{Falcke1996a}
pointed out that this is in principle inconsistent, since the
longitudinal pressure gradient would inevitably lead to some acceleration
of a modestly relativistic jet. As a first-order assumption the
velocity was then assumed to be simply given by a purely
pressure-driven wind in the supersonic regime. This approach had the
advantage that the actual acceleration mechanism of the jet, which is
likely magnetohydrodynamic in origin, could be treated as a black box.

Simulations of the actual acceleration process are actually very difficult and time consuming \citep[e.g., ][]{MeierKoidaUchida2001,DeVilliersHawleyKrolik2005}. However, they all start with some initial magnetohydrodynamic collimation regime (here referred to as the ``nozzle''). After passing through the fast magnetosonic point, the
flow eventually becomes over-pressured in a phase where the jet expands more or less freely into the ambient medium. The latter situation is mainly addressed by simulations of pc-scale jets observed with VLBI \citep[e.g.,][]{2008arXiv0811.1143M}.

Since for our simple Sgr A* jet model only the part after the sonic point
was considered, the only main parameter is then the location of the
sonic point and the sound speed. For powerful, relativistic jets the
sonic point is expected to be up to thousands of Schwarzschild radii
away from the central engine \citep{MarscherJorstadDArcangelo2008}
while for Sgr~A* a relatively small value, of a few
Schwarzschild radii, seems required by the data
\citep{MarkoffBowerFalcke2007}. The magnetized plasma is here treated
as a single-component fluid with adiabatic index 4/3 -- i.e., in the
relativistic limit of a ``photon gas''. The supersonic jet evolution is then
calculated from the modified, relativistic Euler equation for a freely
expanding jet propagating along the $z$ axis in a cylindrical
coordinate system, which we reproduce here from \citet{Falcke1996a}: }

\begin{equation}\label{euler1}
\gb  n {\partial\over\partial z}\left(\gb{\omega\over n}\right)=-{\partial\over\partial z}P.
\end{equation}
$\omega=m_{\rm p}nc^2+U_{\rm j}+P_{\rm j}$ is the enthalpy density of
the jet, $U_{\rm j}$ is the internal energy density, $n$ is the
particle density, and $P_{\rm j}=(\Gamma-1)U_{\rm j}$ is the pressure
in the jet (all in the local rest frame). With a ``total
equipartition'' assumption one gets $U_{\rm j}\simeq m_{\rm p}nc^2$,
hence $\omega=(1+\Gamma)U_{\rm j}$ and ${\omega/n}=(1+\Gamma)m_{\rm p}
c^2=$const at the sonic point $z=z_{0}$.  In the free jet with conical
shape the energy density evolves as $U_{\rm j}\propto
\left(\gb\right)^{-\Gamma}z^{-2}$. The final equation is then given by
Eq. 2 in \citet{Falcke1996a}.

{\mybf Note that for simplicity this equation lacks a gravitational
  term. This term becomes negligible quickly at larger distances and
  for unbound flows corresponding to typical radio frequencies for
  Sgr~A*. This is clearly a deficiency when discussing the nozzle
  region in detail. In the following we will subsume this effect in
  the nozzle size as a free parameter. }

\begin{figure}
\resizebox{\hsize}{!}{\includegraphics{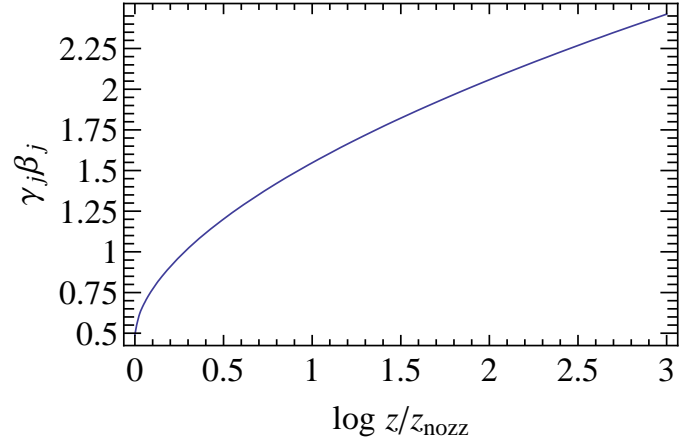}}
\caption{\label{f:sgr-gbplot}
The proper flow speed (Eq.~\ref{gb}) of a pressure driven jet plotted
versus the logarithm of the distance, in units of the nozzle size
$z_{\rm nozz}$, along the z-axis.}
\end{figure}

Previously the solution of the equation was only available numerically
in the code. In in the following we present it in a closed form that
allows testing it against time lag observations. For an adiabatic
index of $\Gamma=4/3$ the solution (see Fig.~\ref{f:sgr-gbplot}) is
given implicitly as

\begin{equation}\label{gb}
\gamma_{\rm j}\beta_{\rm j}=f^\prime\left(8+12\left({4\times3^5\over 5^6\times 7}\right)^{1/6}-28\ln\left({2\over\sqrt{21}}\right)+42\ln{(z)}\right),
\end{equation}
where $f'(y)=x$ is the inverse function of $f$ such that $f(x)=y$, $x=\gamma_{\rm j}\beta_{\rm j}$,  and
\begin{equation}
f(x)=-28 \ln x+{9\over5}42^{2/3}x^{5/3}+42x^2.
\end{equation}
$\gamma_{\rm j}$ and $\beta_{\rm j}c$ are the bulk jet Lorentz
factor and velocity, $z=Z/z_{\rm nozz}$ is the dimensionless length
along the jet axis ($Z$), and $z_{\rm nozz}$ marks the location of the
jet nozzle. The equation thus has a critical point at $z=1$, where
$\gamma_{\rm j}\beta_{\rm j}$ equals the proper sound speed
$\gamma_{\rm s}\beta_{\rm s}=\sqrt{\Gamma(\Gamma-1)(\Gamma+1)^{-1}}$,
and is only valid in the supersonic regime $z>1$.

This relatively simple quasi-analytical description had first been developed for \object{M81*} and was then integrated into the Sgr A* jet papers thereafter. While naturally overly simplified, we retain it here, treating it as a published prediction. {\mybf However, one should not consider this as the only possible solution,   but rather as representative of a broader class of models for modestly relativistic   accelerating jets.}

Using this description we now calculate the time lags based on the
assumption that any flare is essentially due to an increase in the
accretion power. This increased accretion will turn into an increased
outflow rate, based on the ``jet-disk symbiosis''-ansatz of a linear
coupling between inflow and outflow rate
\citep{FalckeBiermann1995}. The increased power and mass flow will
then propagate along the jet essentially with the local flow
speed. Here we ignore the slightly increased acceleration due to the
increased longitudinal pressure gradient in an overdense region, which
would be a second order effect.

{\mybf We note that earlier we have argued that the X-ray flares in
  Sgr A* are not due to a similar increase in accretion, but rather
  due to additional heating or acceleration of the internal particle
  distributions \citep{MarkoffFalckeYuan2001}. This is entirely
  consistent with our approach here, since in the same paper we showed
  that such heating processes only marginally affect the radio flux in
  the optically thick region. Hence simultaneous radio-X-ray flare are
  not necessarily required. Radio flares, however, required an actual
  increase in accretion rate as also argued here. Of course, it is not
  inconceivable that a sudden increase in accretion rate also leads to
  additional heating and particle accreleration in the inner region of
  disk and jet.

  Indeed, recent observations \cite{MarroneBaganoffMorris2008} seem to
  show that there is a rather long lag (on the order of hours) between
  X-ray/IR-flares and radio flares.  This time scale is much longer
  than free-fall or rotational time scales and consistent with viscous
  processes in the accretion flow linking the two types of flares.}

The predicted radio time lags in the jet model are then calculated as

\begin{equation}
\Delta\tau={\Delta \phi D_{\rm GC} \over \sin i \beta_{\rm j}(z) c}\left(1-\beta_{\rm j} \cos i\right).
\end{equation}
$\Delta \phi=\phi_{\rm Sgr\,A*}(\lambda_0)-\phi_{\rm
Sgr A*}(\lambda)$ and $\lambda_0=1.35$ cm is chosen as the
reference wavelength. This formulation recovers the well-known formula
for apparent superluminal motion ($\beta_{\rm app}=\Delta
\phi D/\Delta\tau$), if the implied flares were observed as moving blobs. 

For the dimensionless length we take $z=\phi_{\rm
Sgr\,A*}(\lambda)/\phi_{\rm Sgr\,A*}(\lambda_{\rm nozz})$ with
$\lambda_{\rm nozz}=0.8$mm. The latter represents the next observing
band above the highest currently available VLBI measurements and
corresponds to a projected size of about $4 R_{\rm g}$ ($R_{\rm
g}=GM_{\bullet}/c^2$), for a black hole mass of
$M_\bullet=3.6\times10^6M_{\odot}$. This is also the typical nozzle
size used in spectral fits \citep[e.g.,
][]{FalckeMarkoff2000,MarkoffBowerFalcke2007}.

\begin{figure}
\resizebox{\hsize}{!}{\includegraphics{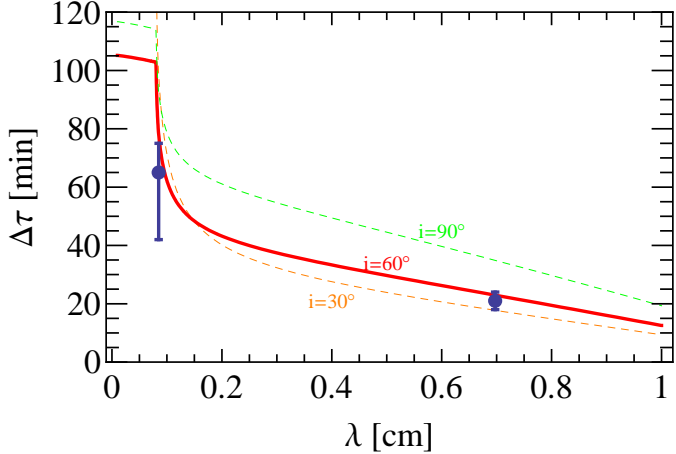}}
\caption{\label{f:sgr-jet-time-delay-vs-lambda}
Expected time lag of Sgr A* versus wavelength relative to $\lambda$1.3
cm (22 GHz) for the observed size-wavelength relation and a flow speed
according to the jet model for three inclination angles. The data
points are the same as in Fig.~\ref{f:sgr-time-delay-vs-lambda}.}
\end{figure}

Figure~\ref{f:sgr-jet-time-delay-vs-lambda} shows the expected time
lag for the measured size and the velocity field of the
pressure-driven jet. The prediction is consistent with the 21 min time
lag found between $\lambda$7 mm and $\lambda$1.3 cm. We stress
that this is based solely on the combination of the observed sizes and
the previously published velocity field for the jet model.

Quite noticeable is the quick rise of the time lag, relative to
$\lambda$1.3 cm, towards shorter wavelengths. The rise comes from the
fact that the jet first needs to accelerate beyond the nozzle, which
yields initially slower flow velocities and accordingly longer time
lags. This should be a characteristic signal of a nozzle, which future
time lag measurements could help to identify. Of course, one has to
bear in mind that the model is overly simplistic and in reality
this feature may look less drastic. In particular, general relativistic
effects will start to play an important role. Also, the exact location
of this kink is very sensitive to the size of the nozzle, which
is a free parameter within a factor of two or so and therefore
difficult to predict. On the other hand, an exact localization of the
kink would effectively constrain the nozzle size.

\begin{figure}
\resizebox{\hsize}{!}{\includegraphics{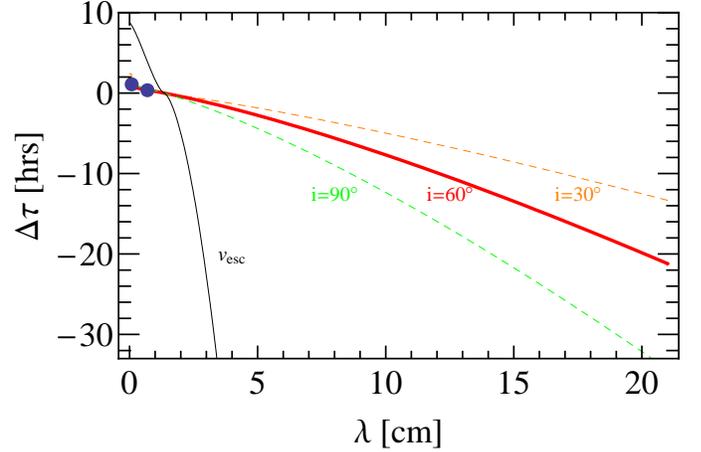}}
\caption{\label{f:sgr-jet-time-delay-vs-lambda-long}Same as Fig.~\ref{f:sgr-jet-time-delay-vs-lambda} 
but extended to longer wavelengths. For reference we also show the
time lags for a marginally bound outflow as in Fig.~\ref{f:sgr-time-delay-vs-lambda}.}
\end{figure}

For future use, we also extrapolate the predicted time lag to longer
wavelengths (Fig.~\ref{f:sgr-jet-time-delay-vs-lambda-long}). One can
see that the lag becomes on the order of a day at cm wavelengths. This may
therefore be difficult to observe, given the limited observability of Sgr A*
in the Northern hemisphere, but would certainly provide crucial
information on the large-scale structure of Sgr A* that is otherwise
impossible to obtain due to the strong scatter broadening.

\begin{figure}
\resizebox{\hsize}{!}{\includegraphics{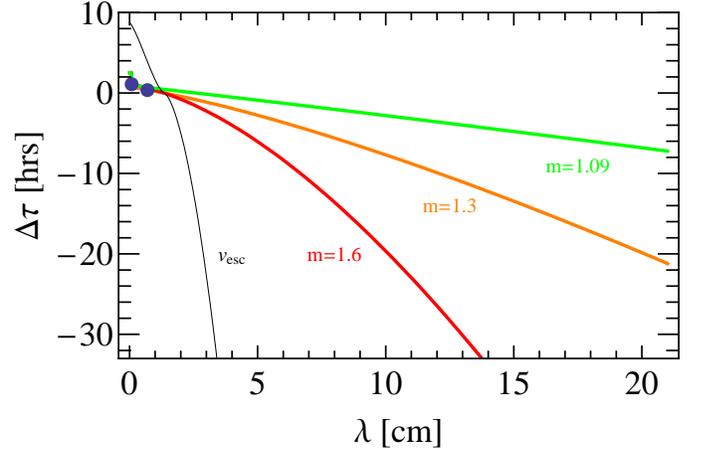}}
\caption{\label{f:sgr-jet-time-delay-vs-lambda-diffsize}Same as Fig.~\ref{f:sgr-jet-time-delay-vs-lambda-long} 
but now we show the time lags for a jet model inclined by $60^\circ$
for the different size-wavelength laws considered. This has a
significant effect at long cm waves.}
\end{figure}

In addition we consider the effect of the range of possible
size-wavelength relations for Sgr A* in
Figure~\ref{f:sgr-jet-time-delay-vs-lambda-diffsize}. Not surprisingly
the time lags do not show much of a difference at short wavelengths,
but differ markedly at long wavelengths.

\section{Summary and discussion}

\label{s:conclusions}

By now Sgr A* is probably the best studied supermassive black hole
with imaging and timing information on scales very close to the event
horizon over a wide range of frequencies. New VLBI measurements, which
we have here revisited, have confirmed theoretical predictions that
Sgr A* has a frequency-dependent photosphere at radio wavelengths,
with sizes scaling roughly as $\lambda^{1.3\pm0.1}$.

The time lag of individual bursts seen at different wavelengths provides a powerful new tool to constrain the physics at work in Sgr A*.  Combining this timing data with the increasingly better intrinsic size measurements obtained with VLBI can significantly constrain flow speeds. The latter is otherwise not measurable with direct imaging due to the extreme scatter broadening in the Galactic center.

The well-established lag of $\sim 20$ min between
$\lambda7$mm and $\lambda1.3$ cm found by
\citet{Yusef-ZadehWardleRoberts2006b}, together with the intrinsic size
difference of $\sim27$ light minutes at these wavelengths, already suggests
that the radio emitting plasma is unbound and flows out with mildly
relativistic speeds.

This is in contrast with the conclusions by
\citet{Yusef-ZadehWardleRoberts2006b} who derive a subsonic and
sub-relativistic outflow from the same time lag data. However, the
authors base their conclusions solely on a fit of their light curve to
a simple \citet{vanderLaan1966} model without ever considering the
VLBI size measurements.  As mentioned by these authors, the van der
Laan model describes the adiabatic expansion of a spherical plasma
blob and cannot describe bulk outflow, which we now know is a major
factor in the extragalactic radio sources it was developed for. The
van der Laan model predicts source sizes of 8 Schwarzschild radii
\citep[e.g., Fig.~3 in][]{Yusef-ZadehWardleRoberts2006b} at 22 GHz
while the measured radio size at 22 GHz is 80 Schwarzschild radii. The
variations of $\ga20\%$ of the total flux would thus have to be
produced by a region that is only $\sim$0.1\% of the total volume
compared to the bulk of the plasma. We find this scenario unlikely.

The pressure-driven jet model \citep{FalckeMarkoff2000}, that has been
successfully used to fit size and spectrum of Sgr A* already, is quite
consistent with the combined size and time lag data. No particular
adjusting of parameters is necessary with respect to published jet
models. The main free parameters are the nozzle size and the
inclination angle, for which we have used canonical
values. Coincidentally, both parameters do not affect the $\lambda$7
mm-$\lambda1.3$ cm lag very much. However, as one can see in the
figures, these parameters will become important once
time lags at other wavelengths are available.

The sensitivity to parameters can be turned around to state that
measurements at other wavelengths in the future will provide
invaluable information about the structure of Sgr A*. If time lags can be
found at $\lambda$2 cm or $\lambda$3.5 cm, this would constrain
inclination angle and the size-wavelength relation much better and
even more clearly distinguish between models.

Moreover, time lags at $\lambda$3 mm and $\lambda$1 mm could start to
show evidence for the acceleration region of the outflow (``the
nozzle''), which would be a unique diagnostic for jet and accretion
physics.  However, here one has to caution that our simple analytic
treatment naturally breaks down near the nozzle region and in the
vicinity of the black hole. More sophisticated numerical
magnetohydrodynamic calculations \citep[e.g.,][]{HawleyKrolik2006} are
clearly needed to understand the submm-wave emission. Also, since the
jet likely originates from an inflow somewhere close to the nozzle
region, this inflow could also contribute to the submm-bump emission
\citep[e.g.,][]{YuanMarkoffFalcke2002}. These additional emission
components might eventually decrease the coherence of outbursts across
wavelength in the submm \& mm-wave region.

So far we only have a single reliable time lag from $\lambda7$ mm to $\lambda$1.3 cm on which to base conclusions, with corroborating evidence from a tentative $\lambda$ 0.8 mm to $\lambda$ 1.3 cm lag.  Therefore further time lag measurements at these and other wavelengths and VLBI closure amplitude measurements at long wavelengths are highly desirable. The former is challenging since the expected time scales at longer wavelengths are on the order of a typical observing run. The latter is challenging because typical VLBI arrays resolve out Sgr A* at long wavelengths and long baselines.  Nonetheless, these observations are certainly worth the effort. They promise a wealth of detailed information about how plasma behaves very close to the event horizon of a supermassive black hole. {\mybf Such coordinated multi-wavelength studies will eventually allow a   range of more sophisticated models to be tested in great detail.}

\bibliographystyle{aa} \bibliography{hfrefs}
\end{document}